\documentclass{article}

\usepackage[latin1]{inputenc}
\usepackage{amsfonts}
\usepackage{amsmath}
\usepackage{amssymb}
\usepackage[american]{babel}
\usepackage{graphicx}
\usepackage{subfigure}
\usepackage{multirow}
\usepackage[small]{caption}
\usepackage[affil-it]{authblk}

\def\lbra{\langle}
\newcommand{\bra}[1]{{\left\langle{#1}\right\vert}}
\newcommand{\ket}[1]{{\left\vert{#1}\right\rangle}}

\newcommand{\scalar}[2]{\langle#1|#2\rangle}
\newcommand{\bracket}[2]{|#1\rangle\langle#2|}
\newcommand{\comment}[1]{}

\begin{document}

\title{ \Large{\textsc{\bf Spatial Search on Grids with Minimum Memory}}}%
\author{Andris Ambainis$^1$, Renato Portugal$^2$, Nikolay Nahimov$^1$}
\affil{$^1$ Faculty of Computing, University of Latvia,\\ Raina bulv.~19, Riga, LV-1586, Latvia}
\affil{$^2$ Laborat\'{o}rio Nacional de Computa\c c\~ao Cient\'{\i}fica, \\ Av. Get\'ulio Vargas, 333, Petr\'opolis, 25651-075, Brazil} 
\date{\today}
\maketitle

\begin{abstract}
We study quantum algorithms for spatial search on finite dimensional grids. 
Patel \textit{et al.}~and Falk have proposed algorithms based on a quantum walk without a coin,
with different operators applied at even and odd steps.
Until now, such algorithms have been studied only using
numerical simulations. In this paper, we present the first rigorous 
analysis for an algorithm of this type, showing
that the optimal number of steps is $O(\sqrt{N\log N})$
and the success probability is $O(1/\log N)$, where $N$ is the number of vertices.
This matches the performance achieved by algorithms that use other forms
of quantum walks.
\end{abstract}

\section{Introduction}

The quantum spatial search problem can be stated as follows. Suppose that one has a graph with $N$ vertices that represent the places that a quantum robot can be and the edges represent the directions that the robot can move among the vertices. Suppose also that one or a subset of vertices is marked. The goal is to find one marked vertex taking the least number of steps, assuming that the robot can move only to neighboring vertices, and each step takes one time unit. 

Benioff~\cite{Ben02} pointed out that a direct application of Grover's search algorithm~\cite{Gro97a} to the quantum spatial search problem on two-dimensional grids of size $\sqrt{N}\times\sqrt{N}$ does not provide a speedup compared to a search performed by a classical random walk, which takes $O(N\log N)$ steps. Aaronson and Ambainis~\cite{AA03} showed that most of quantum speedup can be recovered by using Grover's search together with a ``divide-and-conquer" strategy that splits the grid into several subgrids and searches each of them. Using this method, the problem can be solved in $O(\sqrt{N}\log^2 N)$ steps. 

The use of coined quantum walks~\cite{PortugalBook} to the quantum spatial search problem was introduced by Shenvi \textit{et al.}~\cite{SKW03}, which developed a quantum search algorithm for the hypercube taking $O(\sqrt{N})$ steps providing a quadratic speedup over classical method using random walk. Ambainis \textit{et al.}~(AKR)~\cite{AKR05} used a similar method to build a quantum search algorithm on two-dimensional grids taking $O(\sqrt{N}\log N)$ steps using the method of amplitude amplification. By introducing an extra qubit into the system, Tulsi~\cite{Tul08} was able to improve the time complexity of AKR's algorithm avoiding the use of amplitude amplification.  Ambainis  \textit{et al.}~(ABNOR)~\cite{ABNOR11} also showed how to eliminate the method of amplitude amplification using the AKR's algorithm and performing a post-processing classical search.

\begin{figure}[!ht]
\setcaptionmargin{.5in}
\centering
\includegraphics[width=2.0in]{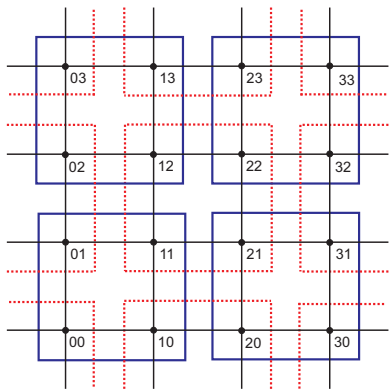}
\caption{Grid tessellation using $2\times 2$ cells. Blue squares (continuous line) represent the even tessellation and red squares (dotted line) represent the odd tessellation.} \label{fig:tessellation2by2}
\end{figure}

Coinless (or staggered) quantum walks for hypercubic lattices were introduced by Patel \textit{et al.}~\cite{PRR05a} by discretizing the Dirac equation used in the staggered lattice fermion formalism. The evolution operator is the product of two unitary operators, which are called \textit{even} and \textit{odd}, and can be obtained from shifted bases via a process of graph tessellation showed in Fig.~\ref{fig:tessellation2by2} for the two dimensional case, which was pointed out by Falk~\cite{Fal13}. Refs.~\cite{PRR05b,PRRA10} also described the use of coinless quantum walks for searching on two-dimensional grids and concluded, using \textit{numerical implementations}, that the search algorithm takes $O(\sqrt{N}\log N)$ steps without using Tulsi's method and $O(\sqrt{N\log N})$ with Tulsi's method. Using a similar algorithm, Falk concluded, also using \textit{numerical implementations}, that the search algorithm takes $O(\sqrt{N})$ steps with constant success probability. 

In this paper we \textit{analytically} prove that a coinless quantum walk using the simplest tessellation (the same one used by Falk) takes $O(\sqrt{N\log N})$ steps to maximize the success probability, which depends on the grid size as $O(1/\log N)$ when there is only one marked vertex. If we use the method of amplitude amplification, the total number of steps is $O(\sqrt{N}\log N)$ in order to achieve a constant success probability $\Theta(1)$.

The structure of this paper is the following: Sec.~\ref{sec:coinless} describes the coinless 
quantum walk model on two-dimensional grids. Sec.~\ref{sec:analysis} describes the general 
structure of the search algorithm, states two claims, and describes the
algebraic manipulation necessary to prove the claims and to find
the number of steps. Sec.~\ref{sec:number} describe the calculation of
the number of steps that
optimize the success probability. Sec.~\ref{sec:norm} describes the
calculation of the norm of the main eigenvector of the evolution
operator, which is used in the analysis of the algorithm.
Secs.~\ref{sec:claim1} and~\ref{sec:claim2} prove the claims. 
In Sec.~\ref{sec:conclusions}, we draw our conclusions and
discuss possible extensions of this work.

\section{Coinless Quantum Walks on Two-Dimensional Grids}\label{sec:coinless}

Consider a two-dimensional grid with $N$ vertices having a torus-like boundary
conditions and the associated Hilbert space ${\cal H}^N$. 
We assume that $N$ is a perfect square and $\sqrt{N}$ is even. 
Define the set of orthonormal vectors
\begin{eqnarray}
	\ket{u^\textrm{even}_{x y}} &=&  \sum_{x',y'=0}^1 a_{x',y'}\ket{2x + x',2y + y'}, \\
	\ket{u^\textrm{odd}_{x y}} &=&  \sum_{x',y'=0}^1 b_{x',y'}\ket{2x + x'+ 1,2y + y' + 1}, 
\end{eqnarray}
which are based in Fig.~\ref{fig:tessellation2by2}.
We address the case $a_{x',y'}=b_{x',y'}=1/2$. The projectors that project into the subspace spanned by $\ket{u^\textrm{e}_{x y}}$ and  $\ket{u^\textrm{o}_{x y}}$ respectively are
\begin{eqnarray}
	\Pi_\textrm{e} &=& \sum_{x,y=0}^{\frac{1}{2}\sqrt{N}-1} \bracket{u^\textrm{e}_{x y}}{u^\textrm{e}_{x y}}, \\
	\Pi_\textrm{o} &=& \sum_{x,y=0}^{\frac{1}{2}\sqrt{N}-1} \bracket{u^\textrm{o}_{x y}}{u^\textrm{o}_{x y}}. 
\end{eqnarray}
Define the reflection operators
\begin{eqnarray}
	U_\textrm{e} &=& 2\, \Pi_\textrm{e} - I,\\
	U_\textrm{o} &=& 2\, \Pi_\textrm{o} - I.
\end{eqnarray}
Define the reflection around the marked vertex
\begin{equation}
  U_w \,=\, 2\,\bracket{w}{w} - I.	
\end{equation}
One step of the quantum walk is driven by the real unitary operator
\begin{equation}
  U \,=\, U_\textrm{o} U_w U_\textrm{e} U_w	
\end{equation}
and the initial state is
\begin{equation}
	\ket{\psi_0} \,=\, \frac{1}{\sqrt N} \sum_{x,y=0}^{\sqrt{N}-1} \ket{x,y}.
\end{equation}
The algorithm consists in obtaining state 
$\ket{\psi_{t_f}}=U^{t_f} \ket{\psi_0}$, 
where ${t_f}$ is the number of steps, and performing a measurement in the
computational basis. The result of the measurement is expected to be
the marked vertex.

\section{Analysis of the Algorithm}\label{sec:analysis}

The eigenvalues of $U$ have the form $\exp(\pm i\theta)$, $0\leq \theta \leq \pi$. Among all eigenvalues different from 1, select the eigenvalue with the smallest positive argument. Let us denote this eigenvalue by $\exp(i\alpha)$ and the associated eigenvector by $\ket{\psi}$. Because $U$ is real, $\exp(-i\alpha)$ is also an eigenvalue and associated with eigenvector $\ket{\psi}^*$. Eigenvectors $\ket{\psi}$ and $\ket{\psi}^*$ are orthogonal and complex conjugates (the entries of $\ket{\psi}^*$ are the complex conjugate of the entries of $\ket{\psi}$). 

Define the vectors 
\begin{eqnarray}
  \ket{\beta^+} &=&  \frac{1}{\sqrt{2}\| \ket{\psi}\|}\left(\ket{\psi}+\ket{\psi}^*\right),\label{beta_+}\\
  \ket{\beta^-} &=&  \frac{1}{\sqrt{2}\| \ket{\psi}\|}\left(\ket{\psi}-\ket{\psi}^*\right),\label{beta_-}
\end{eqnarray}
which are orthonormal. We claim that they define a plane in the Hilbert space ${\cal H}^N$, in which the state of algorithm approximately evolves. This statement is based in the following claims:

\

\noindent
\textbf{Claim 1\,} The overlap $\left|\scalar{\psi_0}{\beta^-}\right|$ between the initial state and $ \ket{\beta^-}$ is $\Theta\left(1\right)$.

\

\noindent
\textbf{Claim 2\,} The success probability $p$ is $O\left(\frac{1}{\log N}\right)$.

\

Claim 1 says that if we replace the initial condition $\ket{\psi_0}$ by $\ket{\beta^-}$, the error will not increase when we increase $N$. The calculation of the evolution of the algorithm is simpler when we take $\ket{\beta^-}$ as the intial condition, because $\ket{\beta^-}$ is a linear combination of only two eigenvectors of the evolution operator, while $\ket{\psi_0}$ has overlap with all eigenvectors.

Suppose that $t_f$ is the number of steps fo the algorithm and take $\ket{\beta^-}$ as the initial state, the final state will be
\begin{eqnarray}
  \ket{\psi_f} &=&  \frac{1}{\sqrt{2}}\left(\textrm{e}^{i\alpha t_f}\ket{\psi}-\textrm{e}^{-i\alpha t_f}\ket{\psi}^*\right). \nonumber
\end{eqnarray}
If we take $t_f=\pi/2\alpha$, then $\ket{\psi_f}= i \ket{\beta^+}$, which is orthogonal to $\ket{\beta^-}$. The success probability is 
\begin{equation}
	p=\left|\scalar{w}{\beta^+}\right|^2.
\end{equation}
The success probability decreases when we increase $N$, but Claim 2 states that the
functional dependence is logarithmic. If one uses the method of amplitude amplification,
the overhead to obtain a constant success probability is $O(\sqrt{\log N})$.

Let us define 
\begin{eqnarray}
  U_1 &=& U_\textrm{e} U_w U_\textrm{e} U_w,	  \nonumber\\
  U_2 &=& U_\textrm{o} U_\textrm{e}.  \nonumber
\end{eqnarray}
Using that $U_\textrm{e}^2=I$, we have $U=U_2 U_1$ with $U_1$ acting as follows: 
$U_1\ket{\psi_1}=\textrm{e}^{2\pi i/3} \ket{\psi_1}$, 
$U_1\ket{\psi_2}=\textrm{e}^{-2\pi i/3} \ket{\psi_2}$ and $U_1\ket{\psi}=\ket{\psi}$
if $\ket{\psi}\perp\ket{\psi_1}$, $\ket{\psi}\perp\ket{\psi_2}$.
In our particular case, the vector $\ket{\psi_2}$ is the complex conjugate of $\ket{\psi_1}$
(i.e., all coefficients of $\ket{\psi_2}$ are complex conjugates of the corresponding coefficients of $\ket{\psi_1}$) and we will use this fact (see the appendix for details).

Let $\ket{v_{j, +}}$ and $\ket{v_{j, -}}$ (for $j=1, 2, \ldots$) be the pairs of 
eigenvectors of $U_2$ with eigenvalues $\textrm{e}^{i\theta_j}$ and $\textrm{e}^{-i \theta_j}$ for $\theta_j\neq 0$
(and $\ket{v_{j, +}}$ is a complex conjugate of $\ket{v_{j, -}}$).
We express
\begin{equation}
\label{eq:psi1} \ket{\psi_1} = a \ket{\psi'_1} + \sum_j \left( a_{j, +} \ket{v_{j, +}} + 
a_{j, -} \ket{v_{j, -}} \right) 
\end{equation}
where $\ket{\psi'_1}$ is an eigenvector of $U_2$ with eigenvalue 1.
By taking complex conjugates of all coefficients of vectors on both sides of the
equation, we get  
\begin{equation}
\label{eq:psi2} \ket{\psi_2} = a^* \ket{\psi'_2} + \sum_j \left( a^*_{j, +} \ket{v_{j, -}} + 
a^*_{j, -} \ket{v_{j, +}} \right) 
\end{equation}
where $\ket{\psi'_2}$ is the complex conjugate of $\ket{\psi'_1}$. The above
equations are valid when $\sqrt{N}/2$ is odd, because $U_2$ has no eigenvalue
-1 in this case. 
Let $\ket{\psi}$ be the eigenvector of $U=U_2U_1$ with the eigenvalue $\textrm{e}^{i\alpha}$
with the smallest positive $\alpha$. We multiply $\ket{\psi}$ by a constant
so that $|\lbra \psi_1 \ket{\psi}|^2 + |\lbra \psi_2 \ket{\psi}|^2 = 1$. Then, we
can express
\[ \ket{\psi} = \cos \beta \ket{\psi_1} + x \sin \beta \ket{\psi_2} + \ket{\psi'} \]
where $\ket{\psi'}\perp \ket{\psi_1}, \ket{\psi'}\perp\ket{\psi_2}$ and $|x|=1$.
To simplify the next expressions, we multiply $\ket{\psi}$ by a constant and $x$ 
by another constant so that 
\begin{equation}
\label{eq:1} 
\ket{\psi} = \textrm{e}^{-i \pi/3} \cos \beta \ket{\psi_1} 
+ \textrm{e}^{i \pi/3} x \sin \beta \ket{\psi_2} + \ket{\psi'} .
\end{equation}
Then, 
\[ U_1 \ket{\psi} = \textrm{e}^{i \pi/3} \cos \beta \ket{\psi_1} 
+ \textrm{e}^{-i \pi/3} x \sin \beta \ket{\psi_2} + \ket{\psi'} .\]
Let 
\[ \ket{\varphi} = U_1 \ket{\psi} - \ket{\psi} = 
(\textrm{e}^{i \pi/3} - \textrm{e}^{-i \pi/3}) \cos \beta \ket{\psi_1} + 
(\textrm{e}^{-i \pi/3} - \textrm{e}^{i \pi/3}) x \sin \beta \ket{\psi_2} \]
\begin{equation}
\label{eq:2}= \sqrt{3} i \cos \beta \ket{\psi_1} - \sqrt{3} x i\sin \beta \ket{\psi_2} .
\end{equation}
By writing out $\ket{\psi_1}$ and $\ket{\psi_2}$ in terms of 
eigenvectors of $U_2$, we get 
\[ \ket{\varphi}=\sqrt{3}i \left(\cos \beta \ket{\psi_1} -  x \sin \beta \ket{\psi_2}\right) =
\sqrt{3}i \Big( a  \cos \beta \ket{\psi'_1} - a^* x \sin \beta \ket{\psi'_2} \]
\begin{equation}\label{eq:phinoprime} +
\sum_j  \left( ( a_{j, +} \cos \beta - a^*_{j, -} x \sin \beta) \ket{v_{j, +}} +
( a_{j, -} \cos \beta - a^*_{j, +} x \sin \beta) \ket{v_{j, -}} \Big) \right).
\end{equation}
Let $\ket{\varphi'} = U_1 \ket{\psi} + \ket{\psi}$. Then, we have
\begin{equation}
\label{eq:3} \ket{\psi}=\frac{1}{2}\ket{\varphi'}-\frac{1}{2}\ket{\varphi}, \mbox{~~~}
U_1\ket{\psi} = \frac{1}{2}\ket{\varphi'}+\frac{1}{2}\ket{\varphi}.
\end{equation}
To get $U_2 U_1 \ket{\psi}=\textrm{e}^{i\alpha} \ket{\psi}$, we must have
\[ \ket{\varphi'} = \sqrt{3} \cot \frac{\alpha}{2} \left(
a \cos \beta \ket{\psi'_1} - a^* x \sin \beta \ket{\psi'_2} \right) \]
\[
 +\sqrt{3} \sum_j   \cot \frac{\alpha-\theta_j}{2}   
( a_{j, +} \cos \beta - a^*_{j, -} x \sin \beta) \ket{v_{j, +}} 
\]
\begin{equation}
\label{eq:varphi}+\sqrt{3} \sum_j \cot \frac{\alpha+\theta_j}{2} 
( a_{j, -} \cos \beta - a^*_{j, +} x \sin \beta) \ket{v_{j, -}} .
\end{equation}
Because of equation (\ref{eq:1}), we have 
\[ \lbra \psi_1 \ket{\psi} = \textrm{e}^{-i \pi/3} \cos \beta = \left( \frac{1}{2} - 
\frac{\sqrt{3}}{2} i \right) \cos \beta .\]
By combining this with the first part of (\ref{eq:3}) and (\ref{eq:2}),
we get that $\lbra \psi_1 \ket{\varphi'} = \cos \beta$. 
Similarly, $\lbra \psi_2 \ket{\varphi'} =  x \sin \beta$.
By writing out $\ket{\psi_1}$, $\ket{\psi_2}$ and $\ket{\varphi'}$ 
in terms of eigenvectors of $U_2$, we get 
\[ \lbra \psi_1 \ket{\varphi'} = \sqrt{3} \cot \frac{\alpha}{2} 
(|a|^2 \cos \beta -(a^*)^2 x \sin \beta \lbra \psi'_1 \ket{\psi'_2}) \]
\[ +\sqrt{3} \sum_j  \cot \frac{\alpha-\theta_j}{2} \left( |a_{j, +}|^2 \cos \beta -
a^*_{j, +} a^*_{j, -} x \sin \beta \right) \]  
\begin{equation}
\label{eq:4}
 +\sqrt{3} \sum_j  \cot \frac{\alpha+\theta_j}{2} \left( |a_{j, -}|^2 \cos \beta -
a^*_{j, -} a^*_{j, +} x \sin \beta \right) =  \cos \beta 
\end{equation}
and 
\[ \lbra \psi_2 \ket{\varphi'} = \sqrt{3}\cot \frac{\alpha}{2} 
(-|a|^2 x\sin \beta + a^2 \cos \beta \lbra \psi'_2 \ket{\psi'_1}) \]
\[ +\sqrt{3}\sum_j  \cot \frac{\alpha-\theta_j}{2} \left( a_{j, -} a_{j, +} \cos \beta -
|a_{j, -}|^2 x \sin \beta \right) \]  
\begin{equation}
\label{eq:5}
 +\sqrt{3}\sum_j  \cot \frac{\alpha+\theta_j}{2} \left( a_{j, +} a_{j, -} \cos \beta -
|a_{j, +}|^2 x \sin \beta \right) =  x \sin \beta. 
\end{equation}


\section{Number of Steps}\label{sec:number}

As described in Sec.~\ref{sec:analysis}, 
the number of steps of the algorithm is $\pi/2\alpha$. The determination of the
asymptotic (large $N$) value of parameter $\alpha$ is the main part to
describe the algorithm efficiency. We address this issue in this section.

We take the complex conjugate of both sides of (\ref{eq:4}) and rewrite the resulting equation 
as 
\begin{equation} \label{eq:A11}
	\sqrt{3}A_{11} \cos \beta + \sqrt{3}A_{12} x^* \sin \beta = \cos \beta
\end{equation}
where
\[ A_{11} = |a|^2 \cot \frac{\alpha}{2} + \sum_j |a_{j, +}|^2 \cot \frac{\alpha-\theta_j}{2} + \sum_j |a_{j, -}|^2 \cot \frac{\alpha+\theta_j}{2} ,\]
\[ A_{12} =  -a^2 \lbra \psi'_2 \ket{\psi'_1} \cot \frac{\alpha}{2} - \sum_j a_{j, +} a_{j, -} \left( \cot \frac{\alpha-\theta_j}{2} + \cot \frac{\alpha+\theta_j}{2} \right) .\]
We can show that, for any $\theta\neq 0$, $\sum_{j: \theta_j=\theta} |a_{j, +}|^2 = \sum_{j: \theta_j=\theta} |a_{j, -}|^2$.
Therefore, we can simplify $A_{11}$ to
\[ A_{11} = |a|^2 \cot \frac{\alpha}{2} + \sum_j \frac{|a_{j, +}|^2 + |a_{j, -}|^2}{2} \left( \cot \frac{\alpha-\theta_j}{2} +  \cot \frac{\alpha+\theta_j}{2} \right) .\]
We can also rewrite (\ref{eq:5}) using that $|x|=1$ as 
\begin{equation} \label{eq:A12}
	-\sqrt{3}A_{11} \sin \beta - \sqrt{3}A_{12}  x^* \cos \beta  = \sin \beta.
\end{equation}

For $\alpha$ close to 0, we can use the approximations $\cot x \approx \frac{1}{x}$ for $\cot\frac{\alpha}{2}$ and 
\[ \cot \frac{\alpha-\theta_j}{2} + \cot \frac{\alpha+\theta_j}{2}  \approx -\frac{\alpha}{\sin^2 (\theta_j/2)} .
\]
Notice that using Eq.~(\ref{eq:costheta}) from the appendix we conclude
that the minimum positive value of $\theta_j$ is $4\pi/\sqrt{N}$. We are
going to show that $\alpha\ll \theta_j$ for large $N$. 
Under those approximations, we obtain
\[ A_{11} \approx  \frac{2|a|^2}{\alpha} - \alpha B,\]
\[ A_{12} \approx  -\frac{2a^2 \lbra \psi'_2 \ket{\psi'_1}}{\alpha} + \alpha C .\]
where
\[ B =  \sum_j \frac{1}{2 \sin^2 (\theta_j/2)} (|a_{j, +}|^2 + |a_{j, -}|^2 )  ,\]
\[ C = \sum_j \frac{1}{\sin^2 (\theta_j/2)} a_{j, -} a_{j, +}.\]

By eliminating $A_{12} x^*$ from Eqs.~(\ref{eq:A11}) and~(\ref{eq:A12}), we obtain
\[ \cos\beta = \frac{1}{\sqrt{2}}\left(1+\frac{1}{\sqrt{3}A_{11}}\right)^{\frac{1}{2}}. \]
By multiplying $(-\sin\beta)$ to Eq.~(\ref{eq:A11}) and adding to Eq.~(\ref{eq:A12}) times $\cos\beta$, we obtain
\[A_{11} \sin\,2\beta + A_{12}x^*=0.\]
Using the last expressions we have obtained for $A_{11}$ and $A_{12}$, we get
\[ \alpha^2 = \frac{2|a|^2\sin\,2\beta - 2a^2 \lbra \psi'_2 \ket{\psi'_1} x^*}{B\sin\,2\beta-Cx^*}.\]
The leading term (zeroth order in $N$) in the numerator of $\alpha^2$ is zero
if $x=|a|^2/a^{*\,2} \lbra \psi'_1 \ket{\psi'_2}$. We use this fact to calculate 
the value of $x$.
Using the eigenvectors and eigenvalues of $U_2$ given in the appendix, we obtain
\begin{equation}\label{eq:moduloasquare}
	|a|^2 =\frac{1}{3} + \frac{8}{3N} + O\left(\frac{1}{N^2}\right)
\end{equation}
and
\begin{equation}\label{eq:asquare}
	a^2 \lbra \psi'_2 \ket{\psi'_1}  =\textrm{e}^{\frac{2\pi i}{3}}\left(\frac{1}{3} - \frac{4}{3N}\right)+ O\left(\frac{1}{N^2}\right). 
\end{equation}
Using that $\alpha^2 B \ll |a|^2$ for large $N$, we can consider $A_{11} \approx 2|a|^2/\alpha$
and
\begin{equation}\label{eq:cosbeta}
	 \cos\beta \approx \frac{1}{\sqrt{2}}\left(1+\frac{\sqrt{3}\alpha}{4}\right).
\end{equation}
Similarly, using that $\sin\,2\beta\approx 1$, the first order approximation for $\alpha$ when $N$ is large is
\begin{equation}\label{eq:alpha2}
	\alpha^2 \approx \frac{8}{N(B-Cx^*)}.
\end{equation}
Using the eigenvectors and eigenvalues of $U_2$, we obtain
\[B-Cx^*= \frac{2}{N}\sum_{
\begin{subarray}{c}
{k,l=0}\\
(k,l)\neq (0,0)
\end{subarray}
}^{\frac{\sqrt{N}}{2}-1}{\frac {1}  { 1 - \cos^2 {\tilde k} \cos^2 {\tilde l}  } },
\]
where ${\tilde k}=2\pi k/\sqrt{N}$ and ${\tilde l}=2\pi l/\sqrt{N}$. Converting the double sum to a double integral and using residues (the expression inside the double sum taken as a function of $\tilde k$ and $\tilde l$ in the domain $(0,\pi)$ has four positive poles), we obtain $B-Cx^*=O\left(\log N\right)$.
Using this result, we conclude that 
\[\alpha=O\left(\frac{1}{\sqrt{N\log N}}\right).\]

\section{The Norm of $\ket{\psi}$}\label{sec:norm}

Using Eqs.~(\ref{eq:phinoprime}) and~(\ref{eq:varphi}), we obtain
\[ \ket{\psi} = \frac{\sqrt{3}}{2} \left(\cot \frac{\alpha}{2}-i\right) \left(
a \cos \beta \ket{\psi'_1} - a^* x \sin \beta \ket{\psi'_2} \right) \]
\[
 +\frac{\sqrt{3}}{2} \sum_j   \left(\cot \frac{\alpha-\theta_j}{2}-i\right)   
( a_{j, +} \cos \beta - a^*_{j, -} x \sin \beta) \ket{v_{j, +}} 
\]
\begin{equation}
\label{eq:psifinal}+\frac{\sqrt{3}}{2} \sum_j \left(\cot \frac{\alpha+\theta_j}{2}-i\right) 
( a_{j, -} \cos \beta - a^*_{j, +} x \sin \beta) \ket{v_{j, -}}. 
\end{equation}
By employing the approximation for small $\alpha$
\[\cot^2 \frac{\alpha\pm\theta_j}{2}+1 \approx \frac{1}{\sin^2\frac{\theta_j}{2}}\mp\frac{2\sin \theta_j\, \alpha}{\left(1-\cos \theta_j \right)^2}.
\]
we obtain
\[\lbra \psi \ket{\psi}\approx \frac{3}{4}\left(|a|^2 - 
a^2 \lbra \psi'_2 \ket{\psi'_1} x^* \sin 2\beta\right)\left(\cot^2 \frac{\alpha}{2}+1\right) 
\]
\[ + \frac{3}{4}\sum_j\frac{1}{
\sin\frac{\theta_j}{2}}\left(|a_{j,+}|^2+|a_{j,-}|^2-2\Re(a_{j,-}a_{j,+}x^*)\sin 2\beta\right).
\]
Using that $\sin 2\beta\approx 1$, Eqs.~(\ref{eq:moduloasquare}) and~(\ref{eq:asquare}), we obtain
\[\lbra \psi \ket{\psi}\approx \frac{12}{N\alpha^2}+\frac{3}{2}\left(B-Cx^*\right). 
\]
Using Eq.~(\ref{eq:alpha2}), we get
\begin{equation}\label{eq:psipsi}
	\lbra \psi \ket{\psi}\approx \frac{24}{N\alpha^2}.
\end{equation}
Therefore, $\| \ket{\psi}\| = O\left(\sqrt{\log N}\right)$.

\section{Proof of Claim 1}\label{sec:claim1}

Let $\ket{\psi_0}$ be the normalized uniform vector (initial condition
of the algorithm). 
We know that $\lbra \psi_0 \ket{v_{j,\pm}}=0$, then for small $\alpha$
\[ \lbra \psi_0 \ket{\psi} \approx \frac{\sqrt{3}}{\sqrt{2}\alpha} \left(
a \lbra \psi_0\ket{\psi'_1} - a^* \lbra \psi_0 \ket{\psi'_2} x \right).
\]
Using Eqs.~(\ref{eq:psi1}) and~(\ref{eq:psi2}) we conclude that
$a \lbra \psi_0 \ket{\psi'_1}= \lbra \psi_0 \ket{\psi_1}$ and 
$a^* \lbra \psi_0 \ket{\psi'_2}= \lbra \psi_0 \ket{\psi_2}$. By
replacing those values into the last equation and using that
$x=\textrm{e}^{2\pi i/3}$, we obtain
\[ \lbra \psi_0 \ket{\psi} \approx \frac{\sqrt{3}}{\sqrt{N}\alpha} \left(
\sqrt{3} - i \right).
\]
Using Eq.~(\ref{eq:psipsi}), we conclude that
the overlap between the initial condition and the normalized 
vector $\ket{\psi}-\ket{\psi}^*$ is
\[  \frac{|\lbra \psi_0 \ket{\psi} - 
\lbra \psi_0 \ket{\psi}^* |}{\sqrt{2}\| \ket{\psi} \|} = \Theta(1).
\]
The asymptotic overlap in this case is 1/2. This overlap can be improved
by changing the global phase of $\ket{\psi}$. In fact, if we take
 $\textrm{e}^{-\pi i/3}\ket{\psi}$, the asymptotic overlap is 1.

\section{Proof of Claim 2}\label{sec:claim2}

Let $\ket{00}$ be the marked vertex. From Eq.~(\ref{eq:psi1}), we
obtain
\[ a\lbra 00 \ket{\psi_1'}= \lbra 00 \ket{\psi_1}-
\sum_j ( a_{j, +} \lbra 00 \ket{v_{j, +}}   - a_{j, -} \lbra 00 \ket{v_{j, -}}  ).  
\]
A similar equation can be obtained for $a\lbra 00 \ket{\psi_2'}$ 
using Eq.~(\ref{eq:psi2}). By employing those results,
the overlap between the 
marked vertex and vector $\ket{\psi}$ can be written as
\[\lbra 00 \ket{\psi} = \frac{\sqrt{3}}{2}\Big( \left(\cot \frac{\alpha}{2}-i\right) \left(
\cos\beta\, \lbra 00 \ket{\psi_1} - x \sin\beta\, \lbra 00 \ket{\psi_2} \right) \]
\[
 +\sum_j  \left(\cot \frac{\alpha-\theta_j}{2}-\cot \frac{\alpha}{2}\right)   
( a_{j, +}\cos\beta  - a^*_{j, -} x \sin\beta) \lbra 00 \ket{v_{j, +}} 
\]
\begin{equation*}
\label{eq:00psi}+\sum_j  \left(\cot \frac{\alpha+\theta_j}{2}-\cot \frac{\alpha}{2}\right)   
( a_{j, -}\cos\beta  - a^*_{j, +} x \sin\beta) \lbra 00 \ket{v_{j, -}}\Big). 
\end{equation*}
By Taylor expanding $\cot \frac{\alpha\pm \theta_j}{2}$ around $\alpha=0$, using 
 $\cot \frac{\alpha}{2}\approx \frac{2}{\alpha}$, Eq.~(\ref{eq:cosbeta}), and 
discarding terms proportional to $\alpha$,
we obtain
\begin{equation}\label{eq:overlap00psi}
\lbra 00 \ket{\psi} \approx  \frac{5\sqrt{3}\, x^*}{8}+\frac{\sqrt 3}{\sqrt{2}\,\alpha}\left(
\frac{i\, x^*}{\sqrt{2}}- E^-\right)-\frac{3}{4\sqrt{2}}\,E^+
 -\frac{\sqrt{3}}{2\sqrt{2}}\, F, 
\end{equation}
where
\begin{equation*}
E^\pm \,=\, \sum_j \left( ( a_{j, +} \pm a^*_{j, -} x ) \lbra 00 \ket{v_{j, +}} +( a_{j, -} \pm a^*_{j, +} x ) \lbra 00 \ket{v_{j, -}} \right)
\end{equation*}
and
\begin{equation*}
F \,=\, \sum_j \cot \frac{\theta_j}{2} \left( ( a_{j, +} - a^*_{j, -} x ) \lbra 00 \ket{v_{j, +}}
-( a_{j, -} - a^*_{j, +} x ) \lbra 00 \ket{v_{j, -}} \right).
\end{equation*}
By employing  the expressions for
$a_{j, \pm}$ and $\ket{v_{j, \pm}}$ given in the appendix, it
is straightforward to show that
\[  
 E^{-} \,=\, \frac{\sqrt {2}\,(\sqrt 3-i)}{N} \sum_{
\begin{subarray}{c}
{k,l=0}\\
(k,l)\neq (0,0)
\end{subarray}
}^{\frac{\sqrt{N}}{2}-1}
\left(1-\frac{\epsilon\sin({\tilde k}+{\tilde l})}{\sqrt{1-\cos^2 {\tilde k}  \cos^2 {\tilde l}}}\right),
 \]
\[
E^{+} \,=\, -\frac{i}{\sqrt 3}\, E^-  - {\frac {1+i\sqrt {3}}{N\sqrt {6}  }} \sum_{
\begin{subarray}{c}
{k,l=0}\\
(k,l)\neq (0,0)
\end{subarray}
}^{\frac{\sqrt{N}}{2}-1}
 {\frac {\sin 2{\tilde k} \, \sin 2{\tilde l}   }{1- \cos^2 {\tilde k}  \cos^2 {\tilde l}}} ,
\]
\[
F \,=\,  \frac{\sqrt {2}\,(1+i\sqrt {3})}{N} \sum_{
\begin{subarray}{c}
{k,l=0}\\
(k,l)\neq (0,0)
\end{subarray}
}^{\frac{\sqrt{N}}{2}-1}
{\frac {\epsilon \,\sin {\tilde k} \, \sin  {\tilde l} }{1-\cos^2 {\tilde k}  \cos^2 {\tilde l} }},
\]
where $\epsilon$ is the sign of $\cos {\tilde k}\,\cos {\tilde l}$. Calculating the 
double sums, we obtain

\[  
 E^{-} \,=\,  \frac{\sqrt{3}-i}{2\sqrt{2}}\left(1-\frac{4}{N}\right),
 \]
\[
E^{+} \,=\,   -\frac{1+i\sqrt{3}}{2\sqrt{6}}\left(1-\frac{4}{N}\right),
\]
and $F \,=\, 0$. By replacing those results into Eq.~(\ref{eq:overlap00psi}), 
we obtain
\begin{equation}\label{eq:00psi}
\lbra 00 \ket{\psi} \approx  -\frac{\sqrt{3}\,(1+i\sqrt{3})}{4}\left(1+\frac{1}{N}\right) +
{\frac{\sqrt{3}\,(\sqrt{3}-i)}{N\alpha}}. 
\end{equation}
For large $N$, the real part of the overlap  $\lbra 00 \ket{\psi}$ tends to $-\sqrt{3}/4$. 
By using the fact that $\| \ket{\psi}\| = O\left(\sqrt{\log N}\right)$, we conclude
that the modulus of the overlap between the marked vertex
and the normalized vector $\ket{\psi}+\ket{\psi}^*$ is
\[  \frac{|\lbra {00} \ket{\psi} +
\lbra {00} \ket{\psi}^* |}{\sqrt{2}\| \ket{\psi} \|} = O\left(\frac{1}{\sqrt{\log N}}\right).
\]

\section{Conclusions and Discussions}\label{sec:conclusions}

We have analyzed the spatial search problem on two-dimensional grids
using the coinless (or staggered) quantum walk model 
introduced by Patel \textit{et al.}~\cite{PRR05a}.
We obtain the asymptotic (large $N$) number of step of the algorithm and the 
asymptotic success probability. We have used the simplest grid tessellation. 
As described in Fig.~\ref{fig:tessellation2by2}, 
we divide the the grid in $2\times 2$ cells having the even-even points in the lowest
left corner of the cells, which provides the even tessellation. The odd
tessellation is obtained by displacing the even tessellation along the diagonal,
so that odd-odd points are in the lowest
left corner. Each cell in the even tessellation
is associated with a normalized uniform vector in Hilbert space ${\cal H}^N$, 
which span a Hilbert subspace of dimension $N/4$. Non-uniform basis vectors
can be used paying a high price in terms of algebraic
manipulations. The unitary operator $U_\textrm{e}$
is a reflection around this Hilbert subspace. Operator $U_\textrm{o}$ is
defined likewise. The product of those two reflections generates a non trivial
unitary operator which defines one step of the coinless quantum walk. 

The spatial search is driven by a unitary operator that interlaces 
the reflection around the marked vertex $U_w$ and operators 
$U_\textrm{o}$ and $U_\textrm{e}$. Patel \textit{et al.}'~choice~\cite{PRR05b}
is $(U_\textrm{o}U_\textrm{e})^3U_w$ while Falk's choice~\cite{Fal13}
is $U_\textrm{o}U_wU_\textrm{e}U_w$. Our analytical calculations use
the latter one. It is interesting to analyze Patel \textit{et al.}'s
model in order to check their numerical results. Patel \textit{et al.}
briefly discuss the use of the unitary operator 
$(U_\textrm{o}U_\textrm{e})^{t_1}U_w$ for $t_1$ smaller than 3.
It is interesting to analyze the case $t_1=1$, which is the 
simplest one.

We have analytically shown that the optimal number of steps of the
search algorithm is $O(\sqrt{N \log N})$ with a success probability
$O(1/\sqrt{\log N})$ when there is only one marked vertex. 
We also assumed that $\sqrt{N}/2$ is odd to simplify the algebraic 
manipulations. A straightforward application of the 
method of amplitude amplification provides an algorithm that
takes  $O(\sqrt{N} \log N)$ steps with success probability $\Theta(1)$.
Alternative methods can be explored, such as, classical post-processing
search similar to the one proposed by ABNOR~\cite{ABNOR11}. It is also
interesting to use of Tulsi's method~\cite{Tul08}, which is based in
the \textit{abstract search algorithm} described in Ref.~\cite{PortugalBook}.
Notice that the abstract search algorithm and
the coinless search algorithm approximately take place in a
two dimensional subspace of the Hilbert space spanned by the 
initial condition and the marked vertex. It is this fact that
is used for obtaining the analytical results of the algorithms.

\section*{Acknowledgments}
A.A. and N. N. were supported by EU FP7 project QALGO (FET-Proactive scheme) and
European Research Council grant MQC.
R.P.~thanks the warm reception at the University of Latvia and
acknowledges financial support from CNPq.


\section*{Appendix}

In this appendix we calculate the eigenvectors and eigenvalues 
of $U_1=U_\textrm{e} U_w U_\textrm{e} U_w$ and $U_2=U_\textrm{o} U_\textrm{e}$, which play 
an essential role in the determination of the special eigenvector of $U=U_2 U_1$ 
associated with the eigenvalue with
smallest positive argument.

If we suppose that the marked vertex is the origin $\ket{w}=\ket{00}$, 
the characteristic polynomial of $U_1$ is
\begin{equation*}
	 \left( {\lambda}^{2}+\lambda+1 \right)  \left(\lambda - 1\right)^{N-2}.
\end{equation*}
In fact, $U_1^3=I$, which shows that the eigenvalues are 1, $\textrm{e}^{\pm 2\pi i/3}$. 
The eigenvector associated with $\textrm{e}^{ 2\pi i/3}$ is
\begin{equation}
	\ket{\psi_1}=\frac{1}{\sqrt{6}} \left(- i\sqrt{3}\,\ket{00} + \ket{01} + \ket{10} + \ket{11} \right),
\end{equation}
and $\ket{\psi_2}=\ket{\psi_1}^*$ is associated with $\textrm{e}^{-2\pi i/3}$. $U_1$ can be expressed as
\begin{equation*}
	U_1 \,=\, I + \sqrt{3}\left(\textrm{e}^{\frac{5\pi i}{6}}\ket{\psi_1}\bra{\psi_1}+\textrm{e}^{-\frac{5\pi i}{6}}\ket{\psi_2}\bra{\psi_2}\right).
\end{equation*}

To obtain the eigenvectors and eigenvalues of $U_2$ we use a staggered Fourier transform,
which can be introduced in the following form.
Define vectors
\begin{equation*}
	\ket{\Psi_{k\,l}} \,=\, a\, \ket{\psi_{k\,l}^{(0)}} + b\, \ket{\psi_{k\,l}^{(1)}} + c\, \ket{\psi_{k\,l}^{(2)}} + d\, \ket{\psi_{k\,l}^{(3)}},
\end{equation*}
where
\begin{eqnarray}
  \ket{\psi_{k\,l}^{(0)}} &=& \frac{2}{\sqrt N}\sum_{i, j =0}^{\frac{\sqrt{N}}{2}-1} \omega^{2ik+2jl} \ket{2i, 2j}, \nonumber\\
  \ket{\psi_{k\,l}^{(1)}} &=& \frac{2}{\sqrt N}\sum_{i, j =0}^{\frac{\sqrt{N}}{2}-1} \omega^{2ik+(2j+1)l} \ket{2i, 2j+1}, \nonumber\\
  \ket{\psi_{k\,l}^{(2)}} &=& \frac{2}{\sqrt N}\sum_{i, j =0}^{\frac{\sqrt{N}}{2}-1} \omega^{(2i+1)k+2jl} \ket{2i+1, 2j}, \label{eqs:psikl} \nonumber\\
  \ket{\psi_{k\,l}^{(3)}} &=& \frac{2}{\sqrt N}\sum_{i, j =0}^{\frac{\sqrt{N}}{2}-1} \omega^{(2i+1)k+(2j+1)l} \ket{2i+1, 2j+1}, \nonumber
\end{eqnarray}
and $\omega=\textrm{e}^{2\pi i/\sqrt{N}}$ and $a,b,c,d$ are complex numbers. For each $k$ and $l$, $\ket{\Psi_{k\,l}}$ span a Hilbert space ${\cal H}_{k\,l}$ that is invariant under the action of $U_2$.
$U_2$ can be expressed as a reduced $4\times 4$--matrix,
\begin{equation}
U_2^{\textrm{red}} \,=\,
 \left[ \begin {array}{cccc} 
{\frac {\cos  {\tilde k} \, \cos  {\tilde l}  }{{\omega}^{{k}+{l}}}}
&{\frac {\sin  {\tilde k}  \, \cos  {\tilde l}  }{{i\,\omega}^{{k}}}}
&{\frac {\cos  {\tilde k}  \, \sin  {\tilde l}  }{i\,{\omega}^{{l}}}}
&\sin  {\tilde k}  \, \sin  {\tilde l}  
\\\noalign{\medskip}{\frac {\sin  {\tilde k}  \, \cos  {\tilde l}  }{i \, {
\omega}^{{k}}}}
&{\frac { {\omega}^{{l}}\cos  {\tilde k}  \, \cos  {\tilde l} }{{\omega}^{{k
}}}}
&-\sin  {\tilde k}  \, \sin  {\tilde l}  
&i\, {\omega}^{{l}}\cos  {\tilde k}  \, \sin  {\tilde l}  
\\\noalign{\medskip}{\frac {\cos  {\tilde k}  \, \sin  {\tilde l}  }{i \,{
\omega}^{{l}}}}
&-\sin  {\tilde k}  \, \sin  {\tilde l}  
&{\frac { {\omega}^{{k}} \cos  {\tilde k} \, \cos  {\tilde l}   }{{\omega}^{{l
}}}}
&i\, {\omega}^{{k}} \sin  {\tilde k}  \, \cos  {\tilde l}  
\\\noalign{\medskip}\sin  {\tilde k}  \, \sin  {\tilde l} \,\,\, 
&i \, {\omega}^{{l}}\cos  {\tilde k}  \, \sin  {\tilde l} \,\,\, 
&i\,{\omega}^{{k}}\sin  {\tilde k} \, \cos  {\tilde l}  \,\,\, 
& {\omega}^{{k+l}}  \cos  {\tilde k}  \, \cos  {\tilde l} 
\end {array}
 \right] ,
\end{equation}
where ${\tilde k}={\frac {2\pi k}{\sqrt N}}$ and ${\tilde l}={\frac {2\pi l}{
\sqrt N}}$. $U_2^{\textrm{red}}$ can be diagonalized and the eigenvalues and 
eigenvectors of this reduced matrix can be used to obtain the eigenvalues and 
eigenvectors of $U_2$ in the original Hilbert space. The eigenvalues of $U_2^{
\textrm{red}}$ are $1$ and $\textrm{e}^{\pm i\,\theta}$, where
\begin{equation}\label{eq:costheta}
	\cos \, \theta \,=\, 2\,  \cos^{2} {\tilde k} \,
 \cos^{2} {\tilde l} -1.
\end{equation}
Note that $\theta$ depends on $k$, $l$, and $N$. The normalized eigenvectors associated with 
eigenvalue 1 are
\begin{equation} 
\ket{w_{k\,l}^{(0)}}\,=\,\frac{1}{2\,c^+}
\left[ \begin {array}{c} \sin  ({\tilde k}-{\tilde l})  
\\ \noalign{\medskip}\sin  {\tilde l}  -\sin  {\tilde k}  
\\ \noalign{\medskip}\sin  {\tilde l}  -\sin  {\tilde k}  
\\ \noalign{\medskip}\sin ({\tilde k}-{\tilde l}) \end {array} \right] , \,\,\,\,\,\,
\ket{w_{k\,l}^{(1)}}\,=\,\frac{1}{2\,c^-}
\left[ \begin {array}{c} \sin  ({\tilde l}-{\tilde k})  
\\ \noalign{\medskip}\sin  {\tilde k}  +\sin  {\tilde l}  
\\ \noalign{\medskip}-\sin  {\tilde k}  -\sin  {\tilde l}  
\\ \noalign{\medskip}\sin ({\tilde k}-{\tilde l}) \end {array} \right], 
\end{equation}
where $(c^{\pm})^2=({ 1 \pm \cos {\tilde k}  \cos  {\tilde l}})({ 1\mp\cos( {\tilde k}-{\tilde l})})$. When $k=l$, the first eigenvector reduces to
\begin{equation*}
	\ket{w_{k\,k}^{(0)}}\,=\,\frac{1}{\sqrt{2}\,\sqrt{1+\,  
\cos^2  {\tilde k}  }}\left[ \begin {array}{c} 1\\ -\cos {\tilde k} \\ -\cos  {\tilde k} 
\\ 1\end {array} \right] 
\end{equation*}
and the second eigenvector reduces to $\ket{w_{k\,k}^{(1)}}=[0,1/\sqrt 2,-1/\sqrt 2,0]$.
Using the fact that $\scalar{\psi_0}{\psi_{k\,l}^{\pm \pm}}=\frac{1}{2}\delta_{k,0}\delta_{l,0}$, it is straightforward to show that the inner products between the initial condition and all those eigenvectors are zero.  Using the fact that $\scalar{00}{\psi_{k\,l}^{(0)}}=\frac{2}{\sqrt N}$ and  $\scalar{00}{\psi_{k\,l}^{(1)}}=\scalar{00}{\psi_{k\,l}^{(2)}}=\scalar{00}{\psi_{k\,l}^{(3)}}=0$, it is straightforward to calculate the inner products between the target state $\ket{00}$ and those eigenvectors, which are
\begin{eqnarray*}
	\scalar{00}{w_{k\,l}^{(0)}}&=&\frac{\sin({\tilde k}-{\tilde l})}{c^+\sqrt{N}}\\
	\scalar{00}{w_{k\,l}^{(1)}}&=&\frac{\sin({\tilde l}-{\tilde k})}{c^-\sqrt{N}}
\end{eqnarray*}
When $k=l$ the inner product between the target and those eigenvectors are $\scalar{00}{w_{k\,k}^{(0)}}=$ $\sqrt{2}/(\sqrt{N}\sqrt{1+\,\cos^2  {\tilde k}})$ and $\scalar{00}{w_{k\,k}^{(1)}}=0$.

The normalized eigenvectors associated with eigenvalue $\textrm{e}^{i\theta}$ are
\begin{equation}
	\ket{w_{k\,l}^{(2)}}\,=\,\frac{1}{2\,c}\left[ \begin {array}{c} -\epsilon\sqrt {c-\epsilon\sin {\tilde k} \cos {\tilde l} }\,\sqrt {c-\epsilon\cos {\tilde k} \sin {\tilde l}} \,\,
\\ \noalign{\medskip}\sqrt {c-\epsilon\sin {\tilde k} \cos {\tilde l} }\,\sqrt {c+\epsilon\cos {\tilde k} \sin {\tilde l} }
\\ \noalign{\medskip}\sqrt {c+\epsilon\sin {\tilde k} \cos {\tilde l} }\,\sqrt {c-\epsilon\cos {\tilde k} \sin {\tilde l} }
\\ \noalign{\medskip}\epsilon\sqrt {c+\epsilon\sin {\tilde k} \cos {\tilde l} }\,\sqrt {c+\epsilon\cos {\tilde k} \sin {\tilde l} }
\end {array} \right],
\end{equation}
where $c^2=1-\cos^2 {\tilde k}\cos^2 {\tilde l}$ and $\epsilon$ is the sign of $\cos {\tilde k} \cos {\tilde l}$. Note that $c\ge \sin {\tilde k} \cos {\tilde l}$. When $k=l$, they reduce to
\begin{equation*}
	\ket{w_{k\,k}^{(2)}}\,=\,{\frac {1}{2\sqrt{1+ \cos^{2} {\tilde k} }}}
\left[ \begin {array}{c} \cos {\tilde k} -\epsilon\sqrt {1+ \cos^2 {\tilde k}}\\ \noalign{\medskip}1
\\ \noalign{\medskip}1\\ \noalign{\medskip}\cos {\tilde k} + \epsilon
\sqrt {1+ \cos^2 {\tilde k}}\end {array}
 \right].
\end{equation*}
Using the fact that the entries of $\ket{w_{k\,l}^{(2)}}$ are real and $(U_2^{\textrm{red}})^*=M U_2^{\textrm{red}} M$, where
\begin{equation*}
	M\,=\,\left[ \begin {array}{cccc} 0&0&0&1\\ 0&0&1&0
\\0&1&0&0\\ 1&0&0&0\end {array}
 \right], 
\end{equation*}
we show that the eigenvectors associated with eigenvalue $\textrm{e}^{-i\theta}$ are obtained by inverting the lines of the eigenvectors associated with $\textrm{e}^{i\theta}$. Instead of inverting the entries of eigenvector, one can invert the sign of $\epsilon$. Notice that for some values of $k$ and $l$ eigenvalues $\textrm{e}^{\pm i\theta}$ can be 1. Also notice that eigenvalues $\textrm{e}^{\pm i\theta}$ can be $-1$ only if $\sqrt{N}/2$ is even.

The eigenvectors of the full matrix are
\begin{equation}\label{eq:Wkl}
	 \ket{v_{k\,l}^{(\beta)}}\,=\,\sum_{\beta'=0}^3 \, \scalar{\beta'}{w_{k\,l}^{(\beta)}} \, \ket{\psi_{k\,l}^{(\beta')}},
\end{equation}
for $\beta=0,...,3$ and $0\le k,l < \sqrt{N}/2$, and the eigenvalues are ${w_{k\,l}^{(0)}}={w_{k\,l}^{(1)}}=1$, ${w_{k\,l}^{(2)}}=\textrm{e}^{i\theta}$, and  ${w_{k\,l}^{(3)}}=\textrm{e}^{-i\theta}$. The entries of $\ket{w_{k\,l}^{(\beta)}}$ are represented by $\scalar{\gamma}{w_{k\,l}^{(\beta)}}$, $0\le \gamma\le 3$, to avoid confusion with the notation of the eigenvalues. Notice that $\ket{v_{k\,l}^{(2)}}$ and $\ket{v_{k\,l}^{(3)}}$ are not complex conjugate. In order to check the results that depend on Eqs.~(\ref{eq:psi1}) and~(\ref{eq:psi2}), we have to replace $\ket{v_{k\,l}^{(3)}}$ by the complex conjugate of $\ket{v_{k\,l}^{(2)}}$. 

We can decompose $\ket{\psi_1}$ in the eigenbasis of $U_2$ as
\begin{equation}
	\ket{\psi_1} \,=\, \sum_{k, l =0}^{\frac{\sqrt{N}}{2}-1}\sum_{\beta=0}^3 a_{k\,l}^{(\beta)} \ket{v_{k\,l}^{(\beta)}},
\end{equation}
where
\begin{equation}
	a_{k\,l}^{(\beta)} \,=\, \frac{2}{\sqrt{N}} \scalar{w_{k\,l}^{(\beta)}}{\psi_1^{\textrm{red}}}
\end{equation}
and
\begin{equation}
	\ket{\psi_1^\textrm{red}} \,=\, \frac{1}{\sqrt{6}}
\left[ \begin {array}{c} -i\sqrt{3}\\ \noalign{\medskip}
\omega^{-l}\\ \noalign{\medskip}
\omega^{-k}\\ \noalign{\medskip}
\omega^{-(k+l)}\end {array}
\right].
\end{equation}
The details of the calculation of $|a|^2$, which can be obtained from Eq.~(\ref{eq:psi1}), are
\begin{eqnarray*}
	|a|^2 &=&  \sum_{k, l =0}^{\frac{\sqrt{N}}{2}-1} \left( \left|a_{k\,l}^{(0)}\right|^2+ \left|a_{k\,l}^{(1)}\right|^2\right) +  \left|a_{0\,0}^{(2)}\right|^2 +  \left|a_{0\,0}^{(3)}\right|^2\\
	      &=&  \frac{1}{3} + \frac{10}{3N} - \frac{4}{3N}\sum_{k,l=0}^{\frac{\sqrt{N}}{2}-1}
{\frac { \cos{\tilde k} \sin {\tilde k} \cos{\tilde l} \sin {\tilde l}  } 
 { 1 - \cos^2{\tilde k} \cos^2{\tilde l} } } \\
				&=& \frac{1}{3} + \frac{8}{3N} + O\left(\frac{1}{N^2}\right).
\end{eqnarray*}

\end{document}